\documentclass[useAMS,usenatbib]{mn2e}

\usepackage{graphicx,natbib}
\usepackage{latexsym}
\usepackage{amssymb}
\usepackage{amsmath}
\usepackage{subfigure} 
\usepackage{hyperref}
\usepackage[flushleft]{threeparttable}

\title[Streams in the Aquarius stellar haloes]
{Streams in the Aquarius stellar haloes} 

\author[F.        A.       G\'omez       et       al.]{Facundo      A.
  G\'omez$^{1,2}$\thanks{Email:fgomez@msu.edu},   Amina   Helmi$^{3}$,
  Andrew P. Cooper$^{4}$, Carlos S. Frenk$^{5}$, \newauthor
  Julio F. Navarro$^{6}$, Simon D. M. White$^{7}$\\
  $^{1}$ Department of Physics and Astronomy, Michigan State University, East Lansing, MI 48824, USA\\
  $^{2}$ Institute for Cyber-Enabled Research, Michigan State University, East Lansing, MI 48824, USA\\
  $^{3}$ Kapteyn Astronomical Institute, University of Groningen,P.O. Box 800, 9700 AV Groningen, The Netherlands\\
  ${4}$ National Astronomical Observatories, Chinese Academy of
  Sciences, 20A Datun Road, Chaoyang, Beijing 100012, China\\
  ${5}$ Institute for Computational Cosmology, Department of Physics,
  University of  Durham, South Road,  Durham, DH1 3LE, UK\\
  ${6}$  CIfAR Senior  Fellow,  Department of  Physics and  Astronomy,
  University of Victoria, Victoria BC, Canada V8P 5C2\\
 ${7}$ Max-Planck-Institut f{\"u}r Astrophysik, Karl-Schwarzschild-Str. 1, D-85748, Garching, Germany}
\begin{document}

\date{}

\pagerange{\pageref{firstpage}--\pageref{lastpage}} \pubyear{}

\maketitle

\label{firstpage}

\begin{abstract}

  We use the very high resolution, fully cosmological simulations from
  the {\it Aquarius} project,  coupled to a semi-analytical model of
  galaxy  formation, to  study  the phase-space  distribution of  halo
  stars in  ``solar neighbourhood''-like  volumes.  We find  that this
  distribution is  very rich  in substructure in  the form  of stellar
  streams for all five stellar  haloes we have analysed. These streams
  can be easily identified in velocity  space, as well as in spaces of
  pseudo-conserved  quantities  such  as  $E$ vs.   $L_{z}$.   In  our
  best-resolved local volumes, the number of identified streams ranges
  from  $\approx 300$  to 600,  in very  good agreement  with previous
  analytical predictions, even in  the presence of chaotic mixing. The
  fraction of particles linked to (massive) stellar streams in these volumes can
  be as large as $84\%$.  The number of identified streams is found to
  decrease as a power-law with galactocentric  radius. We show that the
  strongest limitation  to the  quantification of substructure  in our
  poorest-resolved local volumes is particle resolution rather than strong
  diffusion due to chaotic mixing.

\end{abstract}

\begin{keywords}
galaxies: formation -- galaxies: kinematics and dynamics -- methods: analytical -- methods: N-body simulations
\end{keywords}

\section{Introduction}

In the last decade the characterization of the phase-space
distribution of stars in the vicinity of the Sun has become a subject
of great interest.  It is now well-established that by
studying the phase-space distribution function of the Milky Way we may
be able to unveil its formation history \citep{FBH,hreview}.

According  to the current  paradigm of  galaxy formation,  the stellar
haloes of large galaxies such as  our own are mostly formed through the
accretion and mergers of  smaller objects \citep{sz78}, rather than by
a  rapid collapse  of  a  large and  pristine  gas cloud  \citep{ELS}.
During the  process of accretion, these systems  are tidally disrupted
and contribute  gas and  stars to the  final object.  Whereas  the gas
component rapidly forgets its  dynamical origin due to its dissipative
nature,  stars should  retain this  information for  much  longer time
scales because their density  in phase-space is conserved. This memory
is especially easily  accessible in the outer regions  of galaxies, or
outer stellar haloes, where the  dynamical mixing time scales are very
long.  A large amount of observed stellar streams in the outer regions
of  the Milky  Way  supports this  picture  for the  formation of  its
stellar halo.  The Sagittarius stream \citep{ibata94,ibata01b} and the
Orphan stream  \citep{belu07} are just  two examples among  many other
overdensities          recently          discovered         \citep[see
also][]{new02,ibata03,yanny03,belu06, grill, else09}.

This formation  model also predicts that for  Milky Way-like galaxies,
the portion of  the accreted stellar halo that  dominates in the solar
neighbourhood  is   dynamically  old.   \citep[e.g.][]{cooper,tiss12}.
Therefore, fossil signatures of the most ancient accretion events that
a galaxy has  experienced should be buried in  their inner regions or,
in the case of the Milky  Way, close to or in the Solar Neighbourhood.
However, due to the very short dynamical time scales characteristic of
these  regions,  debris  from  accretion  events  is  expected  to  be
spatially  well-mixed  and   therefore  rather  difficult  to  detect.
Furthermore,  in  cold dark  matter  models,  dark  matter haloes  are
expected to be strongly triaxial \citep[see e.g.][]{Allgood06,vera10}.
As a consequence, chaotic orbits  may be quite important and result in
much  shorter mixing  time-scales  \citep[see, e.g.,][]{vol08},  which
leads to a phase-space distribution that is effectively smooth. On the
other  hand, the  dissipative  condensation of  baryons  can induce  a
transformation of  the structure of dark  haloes to a  more oblate and
axisymmetric  shape  in  the  inner  regions  while  preserving  their
triaxial shape in the outer parts \citep[][]{deba,abadi,val10,bryan}.

The first attempt  to quantify the amount of  substructure in the form
of stellar streams expected in the Solar Neighbourhood was carried out
by   \citet{hw}.    They   suggested   that   $\sim   300$   -   $500$
kinematically-coherent  substructures  should be  present  in a  local
volume  around the  Sun.  Although  they considered  a  fixed Galactic
potential, their  results were  later confirmed by  \citet{hws03}, who
analysed a full cosmological simulation  of the formation of a cluster
dark matter  halo scaled down to  galaxy size.  They  also showed that
debris from accretion events appeared to mix on time-scales comparable
to those  expected for integrable potentials.   However, their results
were  based on  a single  N-body simulation  and were  limited  by its
numerical resolution (despite the 66 million particles used).

Stellar streams  in the inner regions  of the Galactic  halo cannot be
detected simply  by looking for  overdensities in the  distribution of
stars on the  sky.  Thus, several different spaces  have been proposed
and  explored to  identify  substructure.  Examples  are the  velocity
space  \citep{hw,hwzz99,W11}, metallicity,  colour and  chemical abundance
spaces  \citep[e.g.][]{font,schlau,anto,val13}  or  spaces defined  by
integrals   of  motion  and   their  associated   orbital  frequencies
\citep[e.g.][]{hz00,bj05,knebe05,mcmillan,klement09,morri09,gh10a,gh10b}.
At first sight,  these studies as well as  that by \citet{hws03} might
be  in tension  with  the results  reported  by \citet{val13}.   These
authors found no substructure in  the integral of motion space of halo
stars formed in a  fully cosmological hydrodynamical simulation.  They
attributed  the lack  of  substructure to  the  strong chaotic  mixing
present in  this simulation.  Their  results were based on  samples of
$\approx 10^{4}$ stellar particles distributed across the entire halo.
This likely imposes a strong limitation on the ability to identify and
quantify substructure.  If, as  discussed by \citet{hws03},  there are
300-500 streams  crossing the Solar neighbourhood a  resolution of at
least  a few thousand  stars in  a local  small volume  (of a  few kpc
across) is  required. Hence a  much larger number  across the whole
halo.  Therefore samples of a  total of 10,000 halo stars would really
be too  small to find  or characterize substructure.  We  confirm this
intuition below,  where we show  that substructure is  clearly visible
when the numerical resolution (particle number) is large enough.

Thanks  to  current  surveys  such  as  RAVE  \citep{rave}  and  SEGUE
\citep{segue}, as  well as the forthcoming  astrometric satellite {\it
  Gaia}  \citep{perry},  model   predictions  are  becoming  testable.
Although to date only a  handful of stellar streams have been detected
in  the Solar  Neighbourhood \citep[e.g.][]{hwzz99,klement09,smith09},
with  the full 6-D  phase-space catalogues  that are  already becoming
available  \citep{maarten,bond10,carollo10,zwitter,burnett,beers12} it
may soon be possible to start deciphering the formation history of the
Milky Way.

In this  work we  revise the predictions  made by  \citet{hws03}.  Our
goal is to characterize and quantify the amount of substructure inside
solar neighbourhood like volumes  obtained from the fully cosmological
very high resolution simulation of galactic stellar haloes modelled by
\citet{cooper}.   In  Section~\ref{sec:methods}  we  briefly  describe
these models.  We characterize the phase-space distribution of stellar
particles      in     ``solar      neighbourhood''      spheres     in
Section~\ref{sec:charac}.  In  Section~\ref{sec:quanti} we measure the
number of stellar streams, and  explore the relevance of orbital chaos
for  halo stars  near  the Sun,  as  well as  the  impact of  particle
resolution  on  the   quantification  of  substructure.   Finally,  in
Section~\ref{sec:conclu}  we  summarise  our results.  

Throughout this work we use the term substructure \emph{only} to refer
to substructure in the form of stellar streams.

\begin{table*}
\centering
\begin{minipage}{138mm}
\centering
\caption{Properties  of the five  {\it Aquarius  haloes} used  in this
  work at $z=0$. The first  column labels the simulation. From left to
  right, the columns  give the virial radius of  the dark matter halo,
  $r_{200}$,  the number  of particles  within $r_{200}$,  the maximum
  circular velocity,  $V_{\rm max}$,  the particle mass,  $m_{\rm p}$,
  the  concentration  parameter, $c_{\rm  NFW}$,  the intermediate  to
  major, $b/a^{\it sn}$, and the minor to major, $c/a^{\it sn}$, axial
  ratios computed using  dark matter particles located within  6 to 12
  kpc,  the  total  stellar  halo  mass,  $M_{*}$,  the  corresponding
  half-light radius, $r_{1/2}$, the  total stellar mass of the central
  galaxy  in {\rm GALFORM},  and the  RMS scatter  in the  logarithm  of the
  stellar mass assigned to the individual particles.}
\label{table:aquarius}
\begin{tabular}{@{}llllllllllll} \hline \hline Name & $r_{200}$ & $N_{200}$
  & $V_{\rm max}$ & $m_{\rm p}$ & $c_{{\rm NFW}}$ & $b/a^{\it sn}$ & $c/a^{\it sn}$ & $M_{*}$ &
  $r_{1/2}$ & $M_{\rm gal}$ & $\log_{10}~(m_{*})$ \\
& & [$10^{6}$] & & [$10^{3}$] & & & & [$10^{8}$] & & [$10^{10}$]& RMS \\
\hline A-2  & $246$ & $135$ &$209$ & $13.7$ & $16.2$ & $0.65$ & $0.53$ & $3.8$ & $20$ &  $1.88$ & $1.06$ \\
B-2 & $188$ & $127$ & $158$ & $6.4$ & $9.7$ & $0.46$ & $0.39$ & $5.6$ & $2.3$ & $1.49$ & $1.30$ \\
C-2 & $243$ & $127$ & $222$ & $14.0$ & $15.2$ & $0.55$ & $0.46$ & $3.9$ & $53$ & $7.84$ & $1.22$ \\
D-2 & $243$ & $127$ &$203$ & $14.0$ & $9.4$ & $0.67$ & $0.58$ & $11.1$ & $26$ & $0.72$ & $2.18$ \\
E-2 & $212$ & $124$ & $179$ & $9.6$ & $8.3$ & $0.67$ & $0.46$ & $18.5$ & $1.0$ & $0.45$ & $1.90$ \\
\hline
\end{tabular}
    \begin{tablenotes}
      \small  
      \item Masses are  in M$_{\odot}$,  distances in  kpc and
        velocities  in km  s$^{-1}$.
      \item Note that our  stellar halo masses ($M_{*}$) also includes
        the mass assigned to the bulge component in \citet{cooper}.
    \end{tablenotes}
\end{minipage}
\end{table*}

\section{The Simulations}
\label{sec:methods}

In this work we analyse the suite of high-resolution $N$-body
simulations from the {\it Aquarius Project}
\citep[][]{springel2008a,springel2008b}, coupled with the {\rm
GALFORM} semi-analytic model \citep[][]{cole94,cole00,bower06} as
described by \citet{cooper}.

\subsection{N-body simulations}

\begin{figure}
\includegraphics[width=84mm,clip]{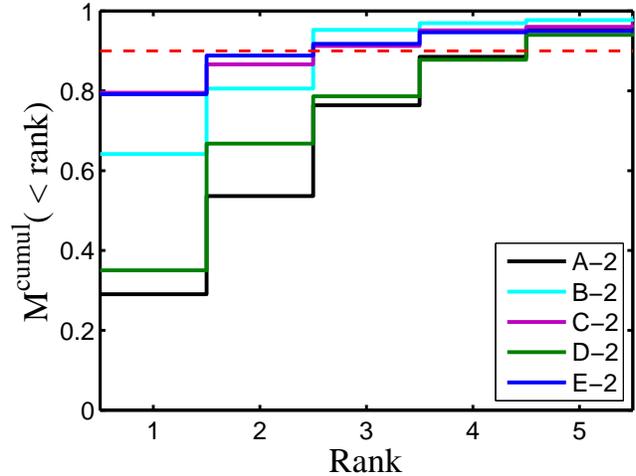}
\caption{Cumulative  stellar mass  fractions obtained  from  our solar
  neighbourhood-like  spheres,  centred at  8  kpc  from the  galactic
  centre.   The step-like cumulative  fractions show  the contribution
  from  the five  most massive  contributors  to each  sphere in  rank
  order.  The most massive contributor is ranked as number 1.  The red
  dashed line shows the $90\%$ level.}
\label{fig:mass_halos}
\end{figure}

The {\it  Aquarius Project} has  simulated the formation of  six Milky
Way-like  dark  matter  haloes   in  a  $\Lambda$CDM  cosmology.   The
simulations were carried out  using the parallel Tree-PM code GADGET-3
\citep[an upgraded version  of GADGET-2,][]{springel2005a}.  Each halo
was   first  identified  in   a  lower   resolution  version   of  the
Millennium-II Simulation  \citep{bk09} which was carried  out within a
periodic box of  side 125 h$^{-1}$ Mpc in  a cosmology with parameters
$\Omega_{m}  = 0.25$,  $\Omega_{\Lambda}= 0.75$,  $\sigma_{8}  = 0.9$,
$n_{s}  =  1$,  and  Hubble  constant  $H_{0}  =  100~h$  km  s$^{-1}$
Mpc$^{-1}$ = 73  km s$^{-1}$ Mpc$^{-1}$.  The haloes  were selected to
have masses comparable  to that of the Milky Way  and to be relatively
isolated  at  $z=0$.   By  applying a  multi-mass  particle  `zoom-in'
technique,  each halo was  re-simulated at  a series  of progressively
higher resolutions.  The  results presented in this work  are based on
the simulations  with the second highest  resolution (Aq-2) available,
with a Plummer  equivalent softening length of 65.8  pc.  In all cases
128  outputs,  starting  from  redshift $z  \approx  45$\footnote{Note
  however that the simulations were started at redshift $z=127$}, were
stored. In each  snapshot, dark matter haloes were  identified using a
Friends-of-Friends  \citep[][]{fof}  algorithm  while  subhaloes  were
identified with  SUBFIND \citep[][]{SUBFIND}.  For  more details about
the      simulations,      we       refer      the      reader      to
\citet{springel2008a,springel2008b}.   The   main  properties  of  the
resulting haloes  are summarised in  Table~\ref{table:aquarius}.  Note
that we exclude  from our analysis the halo Aq-F.  This  is due to its
assembly history; more  than $95\%$ of its accreted  stellar halo mass
comes from a single galaxy that  was accreted very late, at $z \approx
0.7$.

\subsection{Semi-analytic model}
\begin{table}
\centering
\centering
\caption{Properties  of the distribution  of stellar  particles inside
  our  ``solar neighbourhood'' ({\it sn}) spheres  of 2.5  kpc radius.  The first
  column labels the simulation. From left to right, the columns give
  the local stellar density, $\rho_{0}$, the number of most
  significant stellar mass contributors, $N^{\it sn}_{\rm ms}$ (see text) and the three
  components of the velocity ellipsoid.}
\label{table:solar_volume}
\begin{tabular}{@{}llllll}
  \hline \hline \noalign{\smallskip}
  Name & $\rho_{0}/10^4$ & $N_{\rm ms}^{\it sn}$ & $\sigma_{R}$ &
  $\sigma_{\phi}$ & $\sigma_{Z}$ \\
  \small & \small [M$_{\odot}$~kpc$^{-3}$] & & \small [km~s$^{-1}$] & \small [km~s$^{-1}$] & \small [km~s$^{-1}$] \\
  \hline
  A-2 & 1.54 & 5 & 149.4 & 130.9 & 90.5 \\
  B-2 & 15.4 & 3 & 85.3 & 51.1 & 45.5 \\
  C-2 & 2.55 & 3 & 148.2 & 98.1 & 80.7 \\
  D-2 & 8.24 & 5 & 175.1 & 88.5 & 75.7 \\
  E-2 & 3.44 & 3 & 93.8 & 62.3 & 55.0 \\
  \hline
\end{tabular}
\end{table}

To study  the growth and properties of  stellar haloes, \citet{cooper}
coupled the semi-analytic model  {\rm GALFORM} to the {\it Aquarius}
dark   matter   $N$-body   simulations.    The   basic   idea   behind
semi-analytic techniques  is to model the evolution  of the baryonic
components  of  galaxies  through  a  set  of  observationally  and/or
theoretically  motivated  analytic  prescriptions.  Critical  physical
processes  that govern  galaxy formation,  such as  gas  cooling, star
formation feedback,  supernovae, winds of massive stars  and AGN, etc.\,
are taken  into account.  The parameters that  control these processes
were fixed on  the basis of an implementation of  {\rm GALFORM} on the
Millennium  Simulation \citep{springel2005b}, as  this set  of choices
allows one to  successfully reproduce the  properties of galaxies  on large
scales as well as those of the Milky Way \citep{bower06,cooper}.

\begin{figure*}
\centering
\includegraphics[width=175mm,clip]{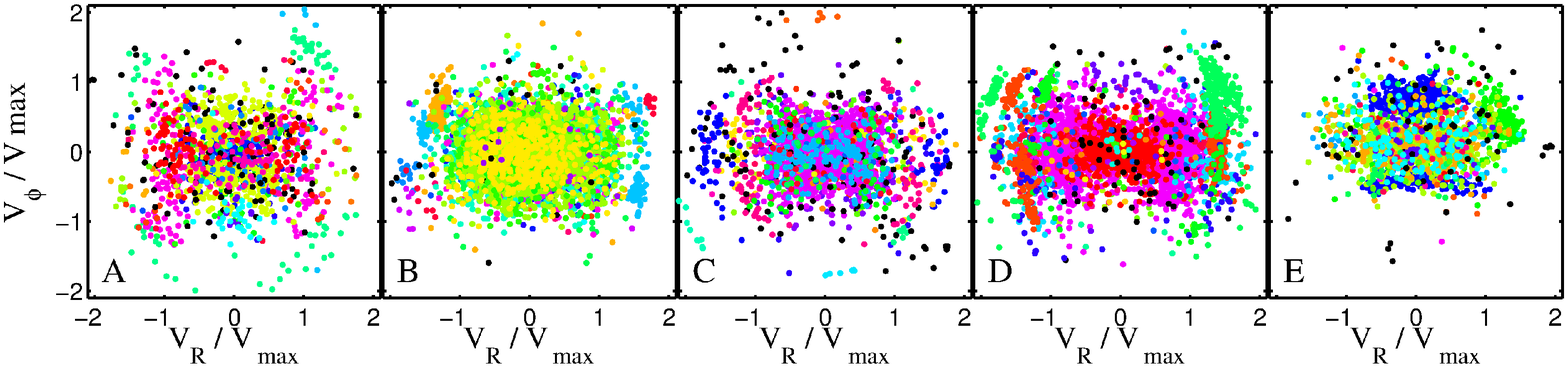}
\\
\includegraphics[width=175mm,clip]{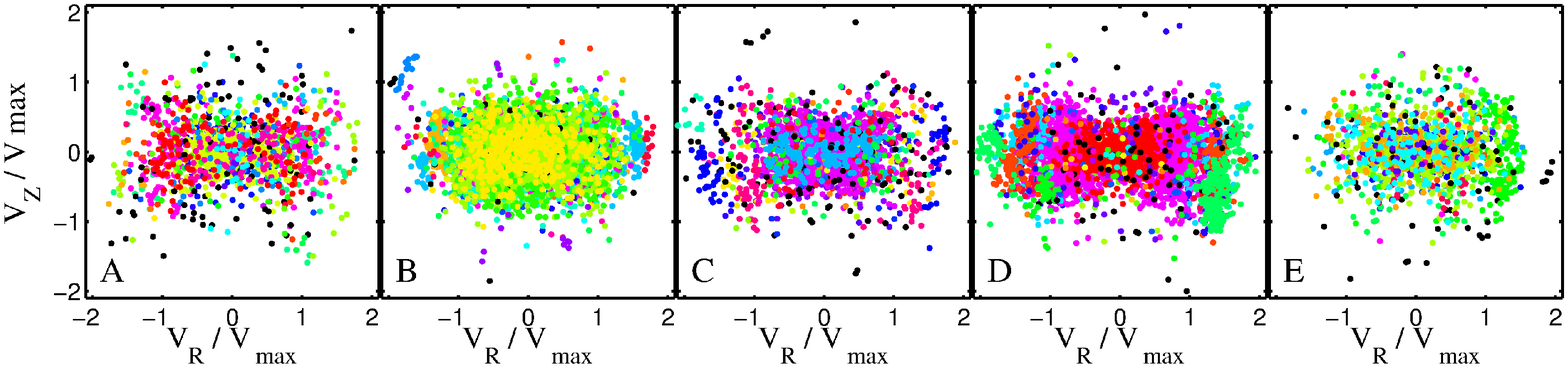}
\caption{Distribution  in  $V_{R}$  vs.  $V_{\phi}$ (top  panels)  and
  $V_{R}$ vs.  $V_{Z}$ (bottom  panels) space of the stellar particles
  located inside a sphere of 2.5 kpc radius at 8 kpc from the galactic
  centre.  The different  colours represent different satellites.  The
  corresponding host  halo is  indicated on the  lower left  corner of
  each panel.}
\label{fig:vela}
\end{figure*}

\begin{figure*}
\centering
\hspace{-0.2cm}
\includegraphics[width=176mm,clip]{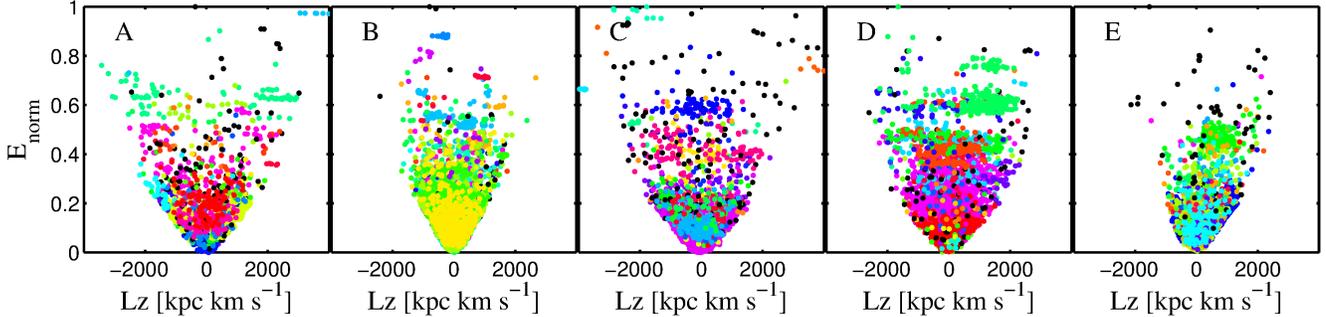}
\caption{Distribution in  $E$ vs.  $L_{z}$  space for the same  set of
  particles  as  shown  in Figure~\ref{fig:vela}.
  The corresponding host  halo is indicated on the  top left corner of
  each panel.}
\label{fig:ener}
\end{figure*}

\begin{figure*}
\centering
\hspace{-0.2cm}
\includegraphics[width=175mm,clip]{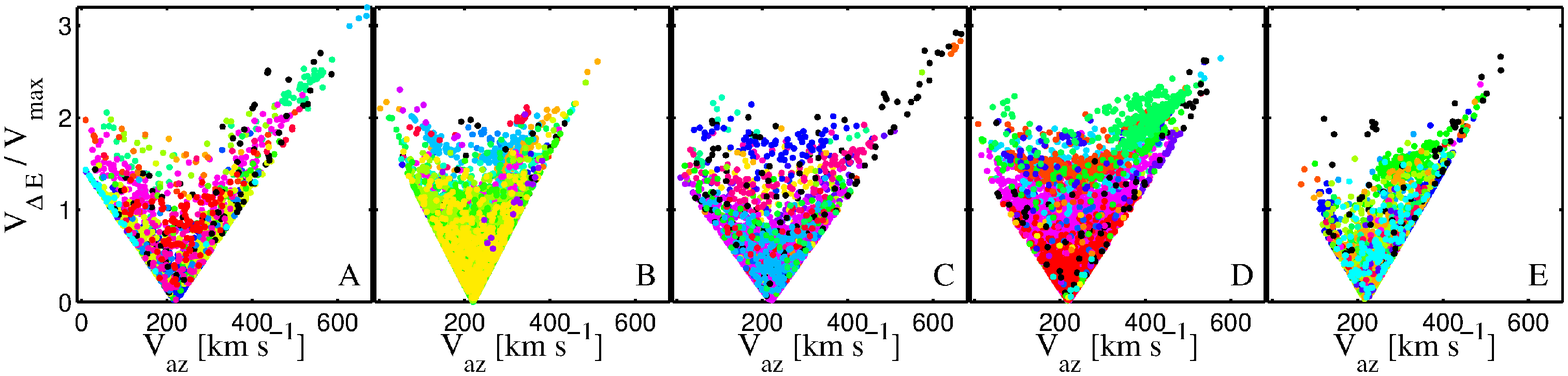}
\\
\includegraphics[width=175mm,clip]{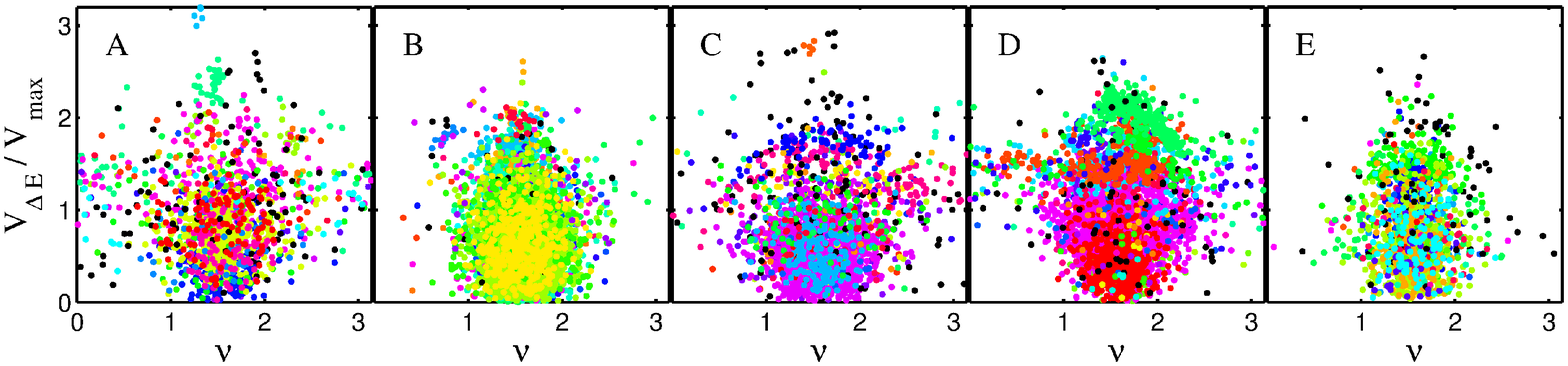}
\caption{Distribution in $V_{\rm az}$ vs.  $V_{\Delta E}$ (top panels)
  and $\nu$ vs.   $V_{\Delta E}$ (bottom panels) space  for the set of
  particles shown in previous figures.  The corresponding host halo is
  indicated on the lower right corner of each panel.}
\label{fig:klm}
\end{figure*}

\subsection{Building up stellar haloes}
\label{sec:build}

Within  the  context  of  CDM,   stellar  haloes  may  be  formed  via
hierarchical  aggregation  of  smaller   objects,  each  one  of  them
imprinting their  own chemical and  dynamical signatures on  the final
halo's  properties  \citep{FBH}.  As  previously  explained, the  {\rm
  GALFORM}  model provides  a  detailed description  of the  composite
stellar  population of  every galaxy  in the  simulation at  any given
time.  However it does not follow the dynamics of these systems fully.
To study  the dynamical properties  of the resulting  galactic stellar
haloes \citet{cooper} assumed that the most strongly bound dark matter
particles  in  progenitor  satellites  could  be  used  to  trace  the
phase-space  evolution  of their  stars.   In  every  snapshot of  the
simulation a  fixed fraction of  the most-bound dark  matter particles
were selected  to trace  any newly formed  stellar population  in each
galaxy in the simulation. This implies that each tagged particle has a
different  final  stellar mass  associated  to  it.   The fraction  of
selected bound particles is set  such that properties of the satellite
population  at  $z=0$, like  their  luminosities, surface  brightness,
half-light radii  and velocity dispersions, are  consistent with those
observed  for  the Milky  Way  and  M31  satellites.  For  a  detailed
description  of this  procedure we  refer the  reader to  Section~3 of
\citet{cooper}.   The  properties  of  the  resulting  stellar  haloes
obtained in each simulation, such as total mass, half-light radius and
the  root-mean-square (RMS) scatter  in the  logarithm of  the stellar
mass   assigned   to   the   individual  particles   are   listed   in
Table~\ref{table:aquarius}.  Note that, although the method introduced
in \citet{cooper}  leads to a successful match  of various observables
regarding the  structure and  characteristics of the  Galactic stellar
halo  and its  satellite population,  the dynamical  evolution  of the
baryonic components of galaxies is much simplified. This likely has an
effect on  e.g.\ the  efficiency of satellite  mass loss due  to tidal
stripping,  or  the satellite's  internal  structural  changes due  to
adiabatic  contraction,   and  possibly  even  on   its  final  radial
distribution. \citep{LY10,RS10,SM11,G13}.

\section{Characterization of Substructure in solar volumes}
\label{sec:charac}

In  this  Section  we  characterize the  phase-space  distribution  of
stellar particles inside a ``solar neighbourhood'' sphere located at 8
kpc  from   the  galactic  centre  of  each   {\it  Aquarius}  stellar
halo. Following  \citet{gh10b}, we chose  for the spheres a  radius of
2.5  kpc  as this  is  approximately  the  distance within  which  the
astrometric satellite {\it Gaia} \citep{perry} will be able to provide
extremely accurate 6D  phase-space measurements for an unprecedentedly
large number of stars. The final configuration of the host dark matter
haloes is,  in all  cases, strongly triaxial.   Therefore, to  allow a
direct  comparison between  the ``solar  neighbourhood'' spheres from
different haloes,  we have rotated each  halo to its  set of principal
axis and placed the sphere along  the direction of the major axis.  In
all  cases, the  ratios  and  directions of  the  principal axis  were
computed using dark matter particles located within 6 to 12 kpc.

Table~\ref{table:solar_volume} shows that  $90\%$ of the total stellar
mass  enclosed  in  these  spheres  comes,  in  all  cases,  from  3-5
significant contributors, $N_{\rm ms}^{\it sn}$.  This is in agreement
with \citet{lh,cooper}, who find that stellar haloes are predominantly
built  from fewer  than 5  satellites  with masses  comparable to  the
brightest   classical   dwarf    spheroidals   of   the   Milky   Way.
Figure~\ref{fig:mass_halos}   shows   the   stellar   mass   fractions
contributed  by the five  most significant  contributors to  the total
stellar mass enclosed in each sphere.

In  Figure~\ref{fig:vela}  we  present  two different  projections  of
velocity space.  The different  colours indicate particles coming from
different  satellites  that  have   contributed  with  at  least  five
particles  (while those from  satellites contributing  fewer particles
are  shown  in  black).   It  is  interesting  to  observe  how  these
distributions in velocity space  vary from halo to halo.  Essentially,
less massive (dark) haloes  have smaller velocity dispersions and thus
the distribution of particles in  velocity space is more compact.  The
values of  the velocity dispersions of these  distributions are listed
in  Table~\ref{table:solar_volume}.  A  comparison with  the estimated
values  of   the  velocity  ellipsoid   of  the  local   stellar  halo
\citep[see][]{chibabeers}   $(\sigma_{R},\sigma_{\phi},\sigma_{Z})   =
(141  \pm 11,  106  \pm  9, 94  \pm  8)$ km  s$^{-1}$  shows that  the
ellipsoids of haloes Aq-A-2,  -C-2 and -D-2 have amplitudes comparable
to those observed for the Milky Way.  Note, however, that the dynamics
of the stellar  particles in these simulations are  only determined by
the underlying dark matter halo  potential.  If the mass associated to
the disc and  the bulge were to be  included, the velocity dispersions
would  be significantly increased,  since we  may relate  the velocity
dispersion $\sigma^{2}$  to the  circular velocity $V_c^2$  and $V_c^2
\propto M(<r) =  M_{\rm disc} + M_{\rm halo}$,  and $M_{\rm disc} \sim
M_{\rm      halo}$     near      the      Sun     \citep{bt}.       In
Table~\ref{table:solar_volume}  we also  show the  local  stellar halo
average densities, $\rho_{0}$.  We  find that, except for halo Aq-B-2,
the values obtained are in reasonable agreement with the estimates for
the     Solar     Neighbourhood,     $\rho_{0}    =     1.5     \times
10^{4}$~M$_{\odot}$~kpc$^{-3}$  \citep[see, e.g.,][]{fuchs}.  However,
one should bear in mind that, if a disc component were to be included,
this should lead  to a contraction of the halo, and  hence to a larger
density,     especially     on     the     galactic     disc     plane
\citep[e.g.,][]{abadi,deba}.  The highest local  density is  found for
halo  Aq-B-2. This  halo has  a low  stellar mass  compared  to haloes
Aq-D-2 and E-2,  but is much more centrally  concentrated than Aq-D-2.
In comparison  to Aq-E-2, which  is also very  centrally concentrated,
Aq-B-2 has  a strongly prolate  shape, which explains its  much higher
value   of  $\rho_{0}$   on  the   major  axis,   where   our  ``solar
neighbourhood'' sphere is placed.

From  Figure~\ref{fig:vela}  we can  also  appreciate  the very  large
amount of  substructure, coming from  many different objects,  that is
present in  these volumes. This  substructure, in the form  of stellar
streams, can be seen as  groups of unicoloured star particles.  Recall
that  these  stellar haloes  are  built  up  in a  fully  cosmological
scenario.  Therefore,  effects such as  violent variation of  the host
potential due to merger  events, and chaotic orbital behaviour induced
by  the  strongly  triaxial  dark  matter  haloes  \citep{vera10}  are
naturally  accounted  for in  these  simulations.  Nonetheless,  these
physical processes have not  been efficient enough to completely erase
the memory  of the origin of  the stellar halo  particles.  Note that,
for this analysis,  we have treated all tagged  particles equally.  As
described in Section~\ref{sec:build},  the stellar-to-total mass ratio
varies from  particle to particle.   Thus, the relative masses  of the
identified streams as  judged by their particle number  could be quite
different  from their  relative stellar  mass.  We  will  explore this
further in the following Section.

Evidence  for this can  also be  seen in  Figure~\ref{fig:ener}, which
shows  the distribution  of  stellar  particles in  the  space of  the
pseudo-conserved quantities $E_{\rm norm}$ and $L_{z}$, where
\begin{equation}
E_{\rm norm}=  \dfrac{E - E_{\min}}{E_{\rm max}  - E_{\rm min}},
\end{equation}
with $E_{\rm max}$ and $E_{\rm min}$  are the energies of the most and
the least  bound stellar particles  inside the volume  under analysis,
respectively.  To compute the energy  we assume a smooth and spherical
representation of the underlying  gravitational potential, as given by
\citet{nfw}
\begin{equation}
\label{NFW_prf}
\Phi(r)=-\frac{GM_{\rm 200}}{r\left[\ln(1+c_{\rm NFW})-c_{\rm
NFW}/(1+c_{\rm NFW})\right]} \\ \ln\left(1+\frac{r}{r_{s}}\right),
\end{equation}
with  values  for  the  parameters  at redshift  $z=0$  as  listed  in
Table~\ref{table:aquarius}.  As  expected, substructure in  this space
is  much  better  defined  than  in  velocity  space  \citep[see  also
e.g.][]{hz00,knebe05,font,gh10a}.   In this projection  of phase-space
streams from  the same satellite tend  to cluster together  and can be
observed as well  defined clumps.  Note however that  within a given
clump substructure  associated with  the various streams  crossing the
``solar neighbourhood'' may be apparent \citep[see, e.g.,][]{gh10a}.

To search for stellar streams in the Solar Neighbourhood without
assuming an underlying Galactic potential, \citet{klement09}
introduced the space of $\nu$, $V_{\Delta E}$ and $V_{\rm az}$, where
\begin{equation}
\begin{array}{lll}
\displaystyle
\label{Klements}
\nu = \arctan\left(\dfrac{V + V_{\rm LSR}}{W}\right), \\
\\
\displaystyle
V_{\Delta{\rm E}} = \sqrt{U^2 + 2(V_{\rm LSR} - V_{\rm az})^2}, \\
\\
\displaystyle
V_{\rm az} = \sqrt{(V + V_{\rm LSR})^2 + W^2}.
\end{array}
\end{equation}
These are based  only on kinematical measurements.  For  a star in the
Galactic plane, $\nu$ would be the angle between the orbital plane and
the direction towards the North Galactic Pole, $V_{\rm az}$ is related
to the  angular momentum and  $V_{\Delta {\rm E}}$  is a measure  of a
star's eccentricity. Under the assumption of a spherical potential and
a  flat rotation  curve, stars  in a  given stellar  stream  should be
distributed  in   a  clump  when  projected  onto   this  space.   The
distribution   of  stellar  particles   located  inside   our  ``solar
neighbourhood'' spheres projected onto  the spaces of $V_{\rm az}$ vs.
$V_{\Delta {\rm E}}$ and $\nu$  vs.  $V_{\Delta {\rm E}}$ are shown in
Figure~\ref{fig:klm}.  Although  perhaps less sharply  defined than in
$E$  vs.   $L_{z}$  space,   substructure  stands  out  in  these  two
projections of phase-space.

\section{Quantification of Substructure in solar volumes}
\label{sec:quanti}

In the previous Section we  found that the phase-space distribution of
stellar particles inside a ``solar neighbourhood'' sphere is very rich
in  substructure in  all  of our  haloes.   We will  now quantify  the
number of  stellar streams crossing the  ``solar neighbourhoods'' of
our {\it  Aquarius} haloes and  characterize their evolution  in time.
Our  goal is to  compare the  results of  this analysis  with previous
studies \citep{hw,hws03} that  have analytically estimated this number
and  concluded that  the  merger history  of  the Milky  Way could  be
recovered  from the  phase-space distribution  of  Solar Neighbourhood
stars.   Furthermore, we  will address  what fraction  of  the stellar
particles that  appear to be  smoothly distributed in  phase-space are
actually in  streams that could  not be resolved  due to the  high but
limited particle resolution of  our $N$-body simulations. We will also
establish the  importance of chaotic  mixing for the  quantification of
substructure.

\subsection{Resolved substructure}
\label{sec:resolved_subs}

To quantify the number of streams  inside our spheres we first need to
specify how to  identify them. Our definition of  a stream must ensure
that particles  in the  same stream share  the same orbital  phase and
progenitor.  Following \citet{hws03}, a stream is identified when {\it
  i)} two or  more particles from the same  parent satellite are found
within  one of  our 2.5  kpc solar  neighbourhood spheres  at redshift
$z=0$ and  {\it ii)} they  share at least  one particle from  the same
parent satellite that has never been separated by more than a distance
$r_{\rm stream}$ from either  of them\footnote{Note that this particle
  does  not need  to be  located within  the  solar neighbourhood-like
  spheres at  $z=0$, as $r_{\rm stream}$  can be larger  than 2.5 kpc.
  Note  further  that  throughout   this  section  when  we  refer  to
  "particles" we mean "particles tagged  with stars'' }.  The value of
$r_{\rm stream}$ is  set to account for the  stellar streams that, due
to  numerical   resolution,  are   poorly  resolved  in   the  ``solar
neighbourhood''. In  this work we  adopt $r_{\rm stream} =  0.8 \times
r_{\rm apo}$ for  each particle, where $r_{\rm apo}$  is its apocenter
at  $z=0$, estimated  from the  outputs of  the  $N$-body simulations.
Varying  the value  of $r_{\rm  stream}$ between  $0.6$ -  $0.9 \times
r_{\rm  apo}$  did not  significantly  affect  our  results.  This  is
required  because  a parent  satellite  can  contribute with  multiple
stellar  streams that  may  cross each  other  in configuration  space
\citep[see,  e.g.,][]{hw}. Note that  using $r_{\rm  apo}$ as  a scale
allows  us  to naturally  adapt  the  algorithm  to different  orbital
configurations.

\begin{figure*}
\centering
\includegraphics[width=170mm,clip]{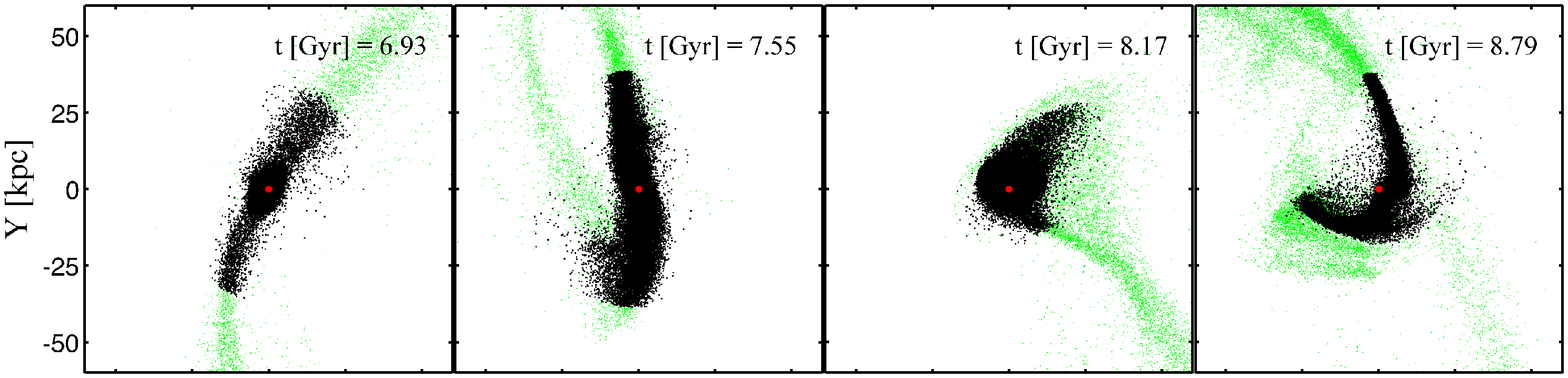}
\\
\includegraphics[width=170mm,clip]{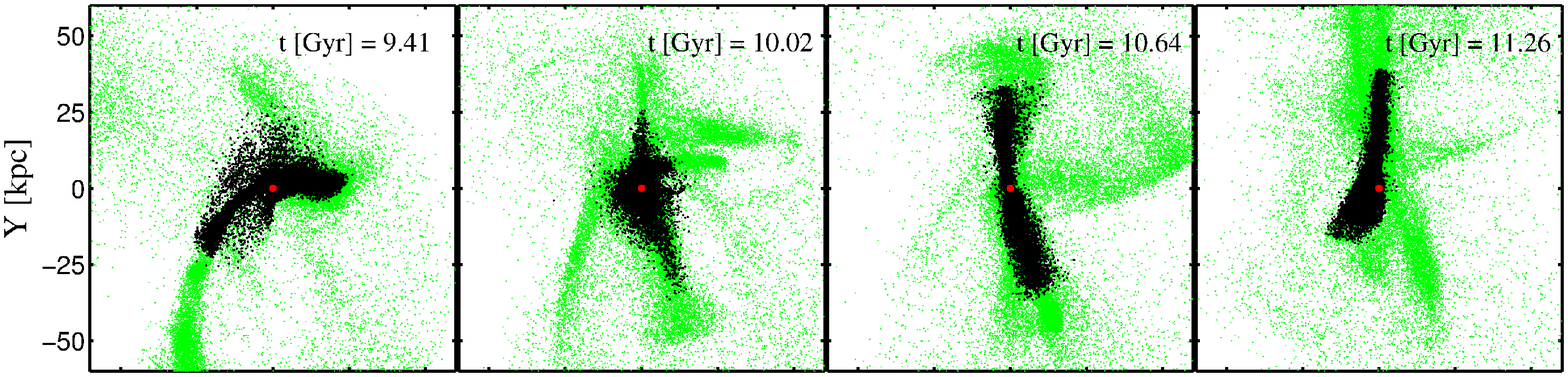}
\\
\includegraphics[width=170mm,clip]{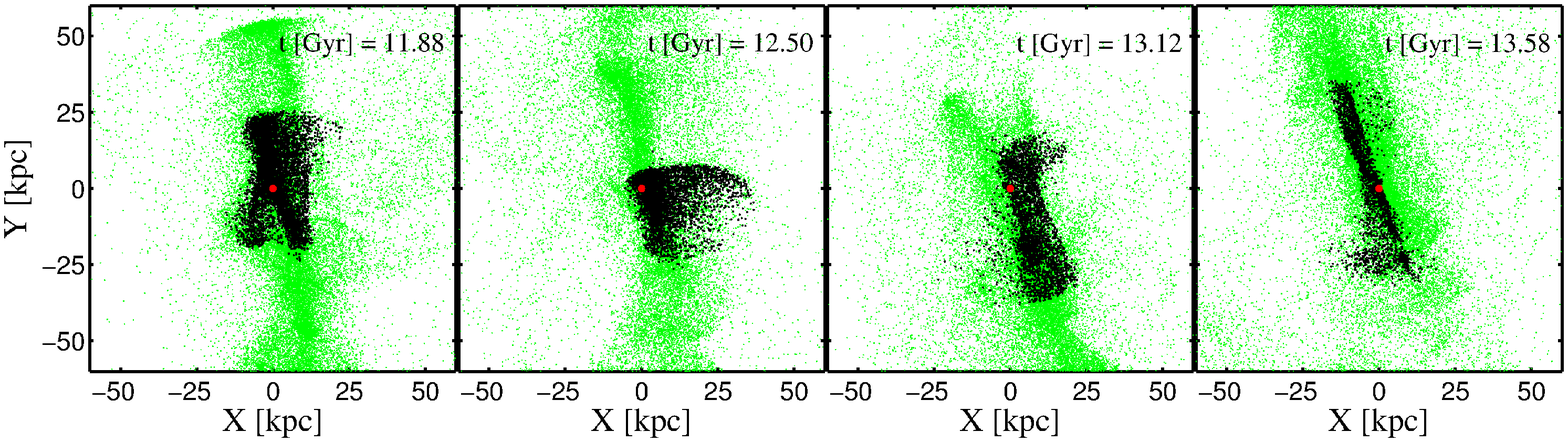}
\caption{Time evolution of a stellar stream that crosses the ``solar
  neighbourhood'' sphere of the {\it Aquarius} halo A-2 at redshift
  $z=0$.  The green dots show all the stellar particles from the
  corresponding parent-satellite.  With black dots we show the
  particles that have always been within $r_{\rm stream}$ from the
  reference particle (red dot), which is indicated with an arrow in
  Figure~\ref{fig:streams_vel}.}
\label{fig:streams_pos_1}
\end{figure*}

In practice, we proceed as follows:
\begin{itemize}
\item For each  particle inside our solar neighbourhood  sphere of 2.5
  kpc radius at redshift $z=0$ we identify its parent satellite.
\item We measure the time $t_{\rm form}$ (prior the time of accretion)
  when the spatial  extent of all particles associated  with this parent
  takes  its  smallest  value  (as measured  by  the  root-mean-square
  dispersion of distance from the centre of mass of the satellite).
\item  At $t=t_{\rm  form}$ we  identify  all particles  located in  a
  sphere of 4 kpc radius,  centred on the selected particle. This
    radius has to be large  enough to initially contain all potential
    neighbouring members of this  given particle.  We have performed various tests and found
    the results to be robust to the sphere's extent.
\item  We follow  these particles  forward  in time  until $z=0$,  and
  discard  those that,  at any  time,  are more  distant than  $r_{\rm
    stream}$ from the selected particle.
\item The  selected particle is  considered to be  in a stream  if, at
  final time, another  particle from the same parent  satellite can be
  found within our solar  neighbourhood sphere and, moreover, they have
  in common at least one of the original neighbouring particles lying
  within their respective $r_{\rm stream}$.
\end{itemize}
\begin{table}
\centering
\begin{minipage}{85mm}
\centering
\caption{Properties of the distribution of particles tagged with stars
  inside ``solar neighbourhood'' spheres  of 2.5 kpc radius located at
  8  kpc  from  the  galactic  centre. The  first  column  labels  the
  simulation. From left to right, the remaining columns give the total
  number of contributing satellites, $N_{\rm sat}^{\it sn}$; the total
  number of star particles, $n_{*}$; the fraction of star particles in
  resolved streams; $f_{\rm  stream}^{n_{*}}$; the fraction of stellar
  mass in resolved streams, $f_{\rm stream}^{m_{*}}$; the total number
  of  streams, $N_{\rm stream}$;  and the  fraction of  streams coming
  from the five most significant stellar mass contributors, $f_{\rm stream}^{5\rm{mm}}$.}
\label{table:streams}
\begin{tabular}{@{}lllllll} \hline \hline \noalign{\smallskip} Name &
  $N_{\rm sat}^{\it sn}$ & $n_{*}$ & $f_{\rm  stream}^{n_{*}}$ & $f_{\rm stream}^{m_{*}}$ & $N_{\rm stream}$  & $f_{\rm stream}^{5\rm{mm}}$ \\
  \hline \hline
  A-2 & $85$ & $1400$ & $20.2\%$ & $31.1\%$ & $83$ & $63\%$ \\
  B-2 & $38$ & $10740$ & $82.0\%$ & $92.4\%$ & $582$ & $74\%$ \\
  C-2 & $105$ & $2334$ & $46.9\%$ & $62.7\%$ & $223$ & $68\%$ \\
  D-2 & $63$ & $3853$ & $68.2\%$ & $85.5\%$ &  $301$ & $65\%$ \\
  E-2 & $53$ & $1957$ & $36.7\%$ & $62.7\%$  & $125$ & $84\%$ \\
  \hline
\end{tabular}
\end{minipage}
\end{table}

\begin{figure*}
\centering
\includegraphics[width=175.5mm,clip]{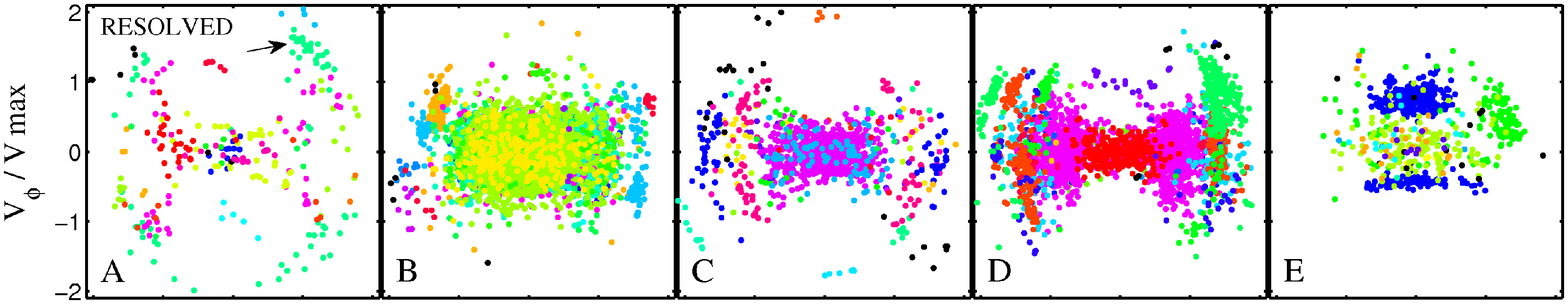}
\\
\includegraphics[width=175mm,clip]{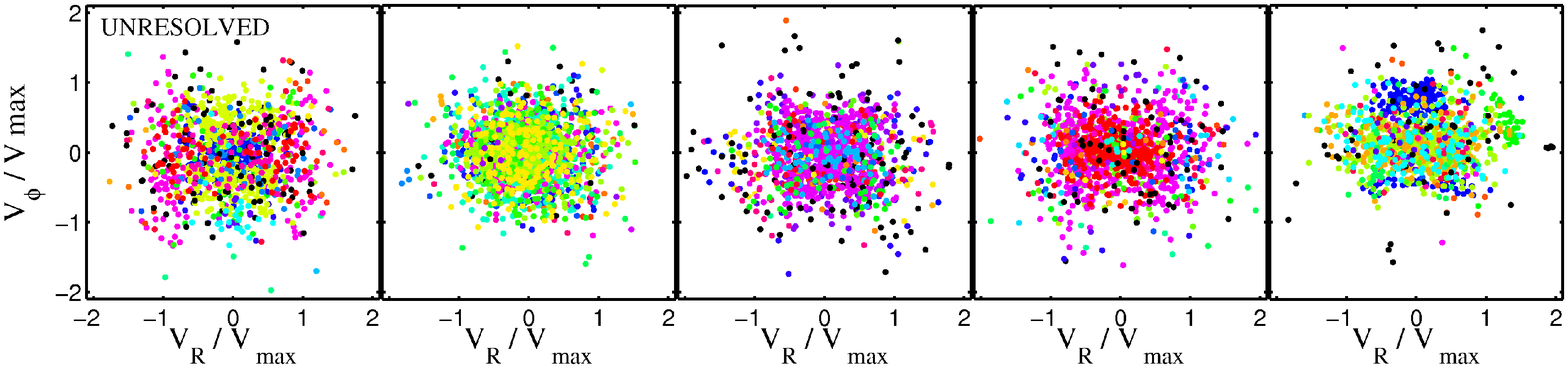}
\caption{Distribution in $V_{R}$ vs.   $V_{\phi}$ space of the stellar
  particles located inside a sphere of 2.5 kpc radius centred at 8 kpc
  from the galactic centre.  The corresponding host  halo is indicated
  on the lower left corner of the top panels. The top  panels show
  the distribution  of particles that  were linked to  stellar streams
  (resolved  component)  whereas   the  bottom  panels  the  remaining
  particles  (unresolved  component).   As  in  Figure~\ref{fig:vela},
  different colours  indicate different satellites.  The  arrow in top
  left  panel of this  figure indicates  the stream  that is  shown in
  Figure~\ref{fig:streams_pos_1}.}
\label{fig:streams_vel}
\end{figure*}

\begin{figure*}
\centering
\includegraphics[width=175.1mm,clip]{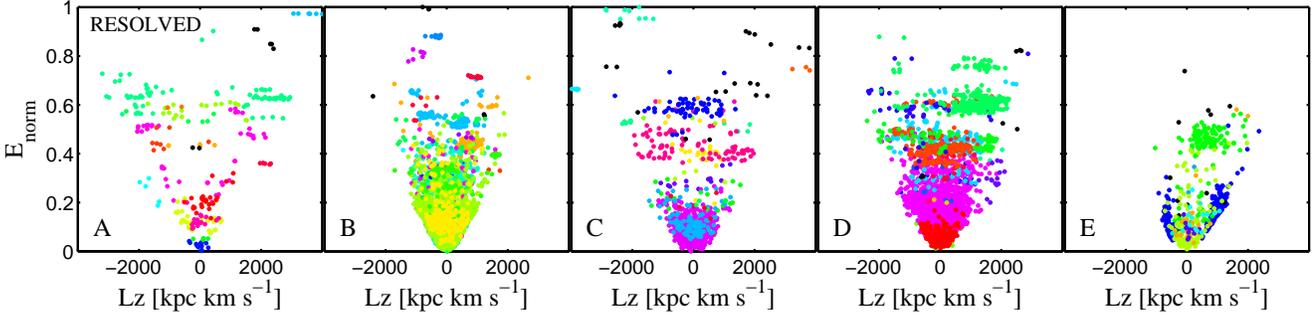}
\caption{Distribution  in  $E$  vs.   $L_{z}$  space  of  the  stellar
  particles that  were linked to stellar  streams (resolved component)
  located inside a sphere of 2.5  kpc radius centred at 8 kpc from the
  galactic centre.   The corresponding host  halo is indicated  on the
  lower left  corner of the top panels.   The colour coding is the same
as in the top panel of Figure~\ref{fig:streams_vel}.}
\label{fig:streams_ener}
\end{figure*}

Figure~\ref{fig:streams_pos_1} shows, as an example, the time
evolution of one of the streams crossing the ``solar neighbourhood''
of the halo Aq-A-2, at $z=0$. The corresponding progenitor is the
second most massive contributor to the stellar halo
and, prior accretion, it had a stellar mass of $\approx 1.25 \times
10^{8}$ M$_{\odot}$.  Its redshift of infall is $z \approx 2.3$ and by
$z = 0$ it has been fully disrupted.  The stream is indicated with an
arrow in Figure~\ref{fig:streams_vel}.  The black dots correspond to
the particles that have always been neighbours (i.e.  within $r_{\rm
  stream}$) of a reference particle in this stream, which is indicated
with a red dot.  This figure clearly shows the full spatial extent of
the stream, which probes regions far beyond the ``solar
neighbourhood'' volume.

The numbers  of streams with at  least two particles  found inside the
``solar  neighbourhood'' spheres located  at 8  kpc from  the galactic
centre  are listed  in Table~\ref{table:streams}.   These  numbers are
well in the  range of the $\sim$ 300 - 500  stellar streams around the
Sun predicted  by the models of  \citet{hw} \citep[see also][]{gh10b}.
Halo Aq-A-2  has the smallest number  of resolved streams,  as well as
the   lowest  fraction   of  particles   associated  with   them  (see
Table~\ref{table:streams}),   and  the  largest   associated  particle
fraction  are  found in  halo  Aq-B-2. Table~\ref{table:streams}  also
shows the total number of  accreted particles found within each sphere
as  well  as the  total  number  of  progenitor satellites  that  have
contributed them.  Note that the  fraction of stellar mass in resolved
streams  is always  larger  than  the fraction  of  star particles  in
resolved streams.  Thus, the star particles with high stellar-to-total
mass ratio, which come from the denser and inner parts of the original
satellites, are more likely to be found in resolved streams.

From this table we see  that disrupted satellites contribute many more
star particles at  8 kpc in Aq-B-2  than in the other haloes,  and as a
result the  streams are much  better resolved. The next  best resolved
``solar neighbourhood''  is found in  Aq-D-2.  For the other  haloes, a
significant fraction  of the particles  have not been  associated with
any  substructure and, therefore  the total  number of  streams (which
would include  those unresolved),  could be significantly  larger.  We
will explore this further in Section~\ref{sec:unresolved}.

In Figure~\ref{fig:streams_vel} we  compare the velocity distributions
of stellar particles  in streams (top panels) to  those not associated
with  any  structure  (bottom  panels)  according  to  our  algorithm.
Different  columns correspond  to distributions  of  stellar particles
inside   ``solar  neighbourhoods''   of  different   haloes.    As  in
Figure~\ref{fig:vela},  the   different  colours  represent  different
satellites.   In addition  to  a very  large  number of  inconspicuous
substructures, we  can see that  the most prominent streams  have been
recovered in  all cases.  Note that  the particles in  streams tend to
populate the wings of the  distribution.  In general, particles in the
core of the velocity distribution  will have been accreted earlier and
tend to  have shorter periods  (and hence to mix  faster).  Therefore,
streams  will be  more difficult  to identify  because of  the limited
resolution.  In addition, and as  explained below, it is possible that
due  to the triaxiality  of the  dark matter  potential some  of these
orbits exhibit  chaotic mixing.  The associated streams  would then be
too convoluted  to be observable.   Note, however, that thanks  to the
much  larger  number  of  star  particles  and  the  smaller  velocity
dispersion (see  Section~\ref{sec:charac}), a significant  fraction of
the  particles in  the  core of  the  distribution for  Aq-B-2 are  in
resolved    streams.     Figure~\ref{fig:streams_ener}    shows    the
distribution of  stellar particles in  streams in integrals  of motion
space.  Substructure  in this space is clearly  visible.  We emphasise
that this is the case even in the presence of violent variation of the
host  potential due  to merger  events and  chaotic  orbital behaviour
induced by the strongly triaxial dark matter haloes.

\begin{figure}
\centering
\includegraphics[width=82mm,clip]{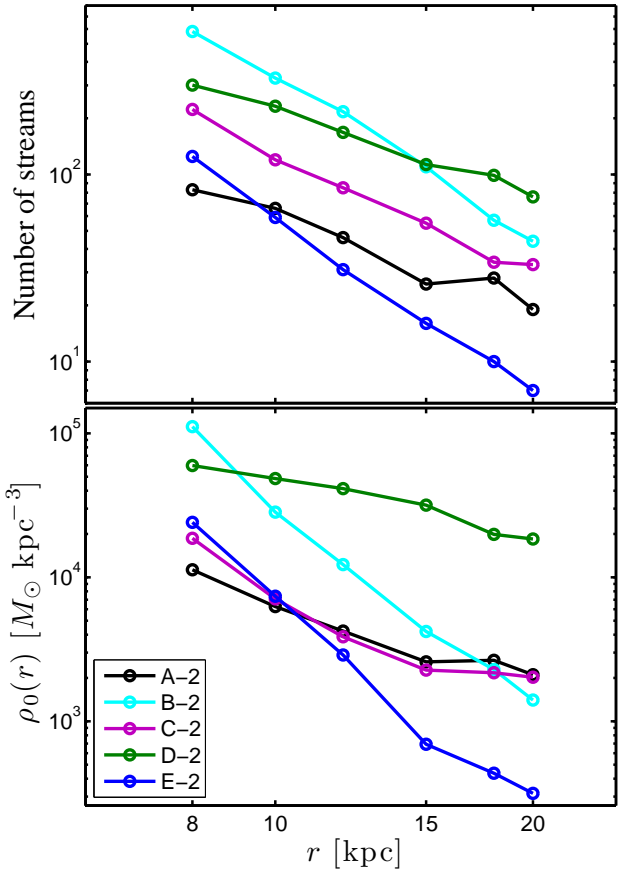}
\caption{Top panel: Number of  streams as a function of galactocentric
  radius  found  within  spheres  of  2.5 kpc  radii.   Bottom  panel:
  Spherically averaged stellar density profiles. The different colours
  indicate different haloes.}
\label{fig:tot_str}
\end{figure}

In the top panel of Figure~\ref{fig:tot_str} we explore how the number
of  streams  found inside  spheres  of 2.5  kpc  radius  changes as  a
function of  galactocentric distance along the direction  of the major
axis.   It is  interesting to  see  that in  all cases  the number  of
streams  decreases as a  power-law with  radius.  Note,  however, that
haloes Aq-B-2 and E-2 have steeper profiles than the other haloes.  As
described by \citet{cooper}, the majority of the stellar mass in these
two cases  comes from one  or two objects  that deposit most  of their
mass  in the  inner regions  of the  haloes. Interestingly,  these two
haloes have the  largest fraction of streams coming  from the five most
significant stellar mass  contributor (see Table \ref{table:streams}).
As  shown in the  bottom panel  of Figure~\ref{fig:tot_str},  the same
behaviours is observed in the  local density profiles of these haloes,
$\rho_{0}(r)$, measured  along the major  axis. Note that, due  to the
triaxial  nature  of the  resulting  haloes,  varying azimuthally  the
location of our  spheres results in local stellar  densities that are,
in general, an  order of magnitude smaller than  the observed value in
the Solar Neighbourhood. For example, at a distance of 8 kpc along the
intermediate  axis,  halo Aq-A-2  has  $\rho_{0}  =  4.9 \times  10^3$
M$_{\odot}$ kpc$^{3}$ and a total number of streams $N_{\rm streams} =
37$. An isodensity contour defined by the value of $\rho_{0}$ at 8 kpc
along  the major axis  places the  solar neighbourhood-like  sphere at
distance  of  $\approx 4$  kpc  on  the  intermediate axis.   At  this
location  $N_{\rm streams}$  raises to  68, indicating  that azimuthal
variations in  the number of  streams mainly reflect changes  in local
stellar density.
 
\subsection{Smooth/Unresolved component}
\label{sec:unresolved}

In addition  to the  stellar particles associated  with streams  with at
least two  particles, our  ``solar neighbourhood'' spheres  contain an
important  number of  stellar particles  that  are not  linked to  any
substructure,  i.e.,   they  appear  to  be   smoothly  distributed  in
phase-space.  The reason for this  could be either of dynamical origin
or, as posited before, simply due to the numerical resolution of the
simulation.   It is  well known that  phase-space regions  of triaxial
dark  matter haloes  may exist  that are  occupied by  chaotic orbits
\citep[e.g.,][]{vol08}.   On such  orbits, stellar  particles  that are
initially  nearby diverge  in  space exponentially  with  time.  As  a
result, substructure that is present in  localised volumes of phase-space left
by  accretion events  is  rapidly  erased. This  process  is known  as
chaotic mixing.   On the  other  hand, initially  nearby particles  in
phase-space on regular orbits diverge in space as a power-law in time.
Therefore, the time-scale in  which these particles fully phase-mix may be
much larger.

Based on the previous discussion, we will now analyse the evolution in
time of the  local (stream) density around the  particles found at the
present day  in each ``solar  neighbourhood'' sphere.  Our goal  is to
assess the  influence of chaotic mixing on  the underlying phase-space
distributions and to estimate the fraction  of particles that, due to the
high but nonetheless limited  numerical resolution of our simulations,
were not associated with any stream.

\begin{figure*}
\centering
\includegraphics[width=85mm,clip]{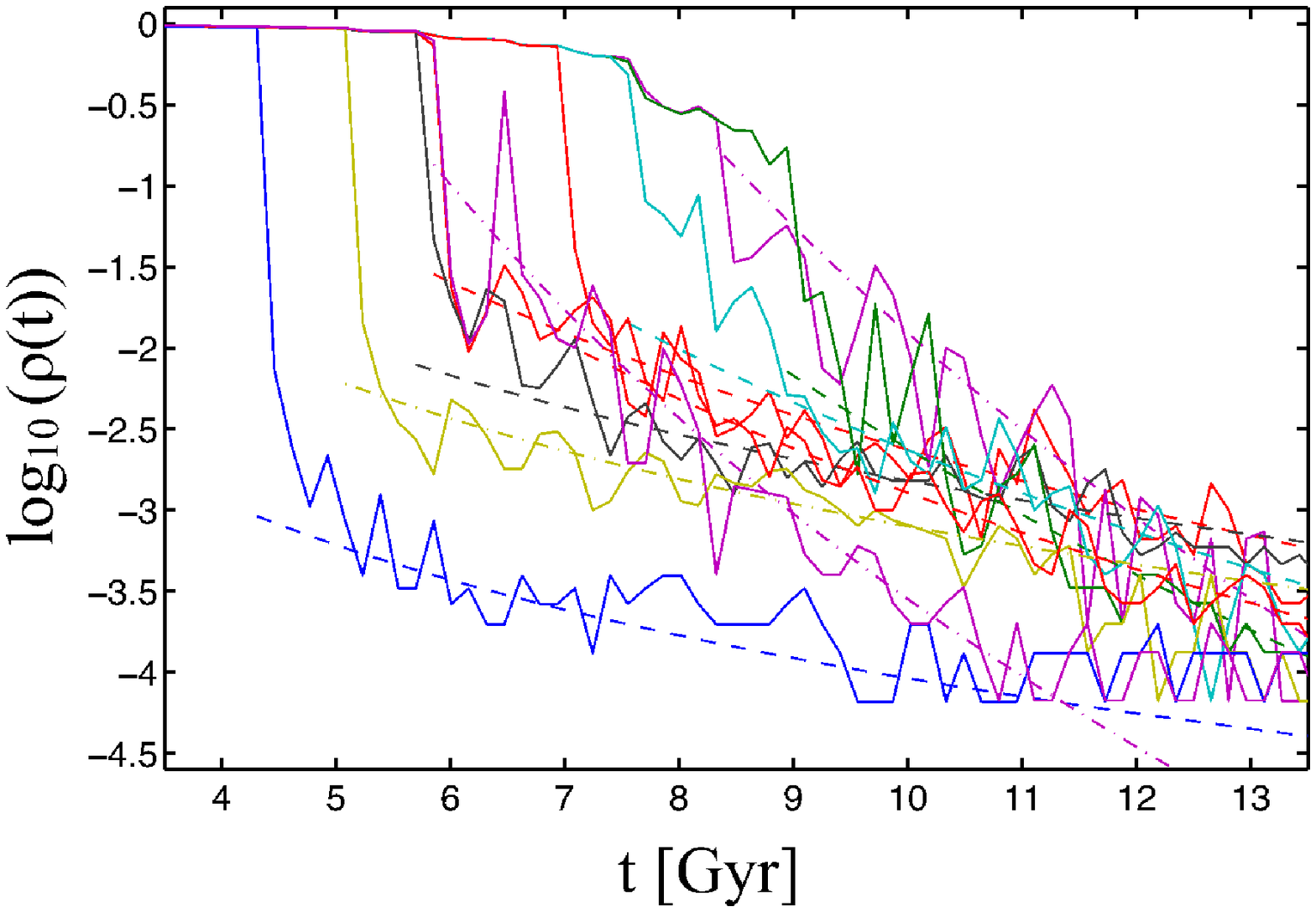}
\includegraphics[width=85mm,clip]{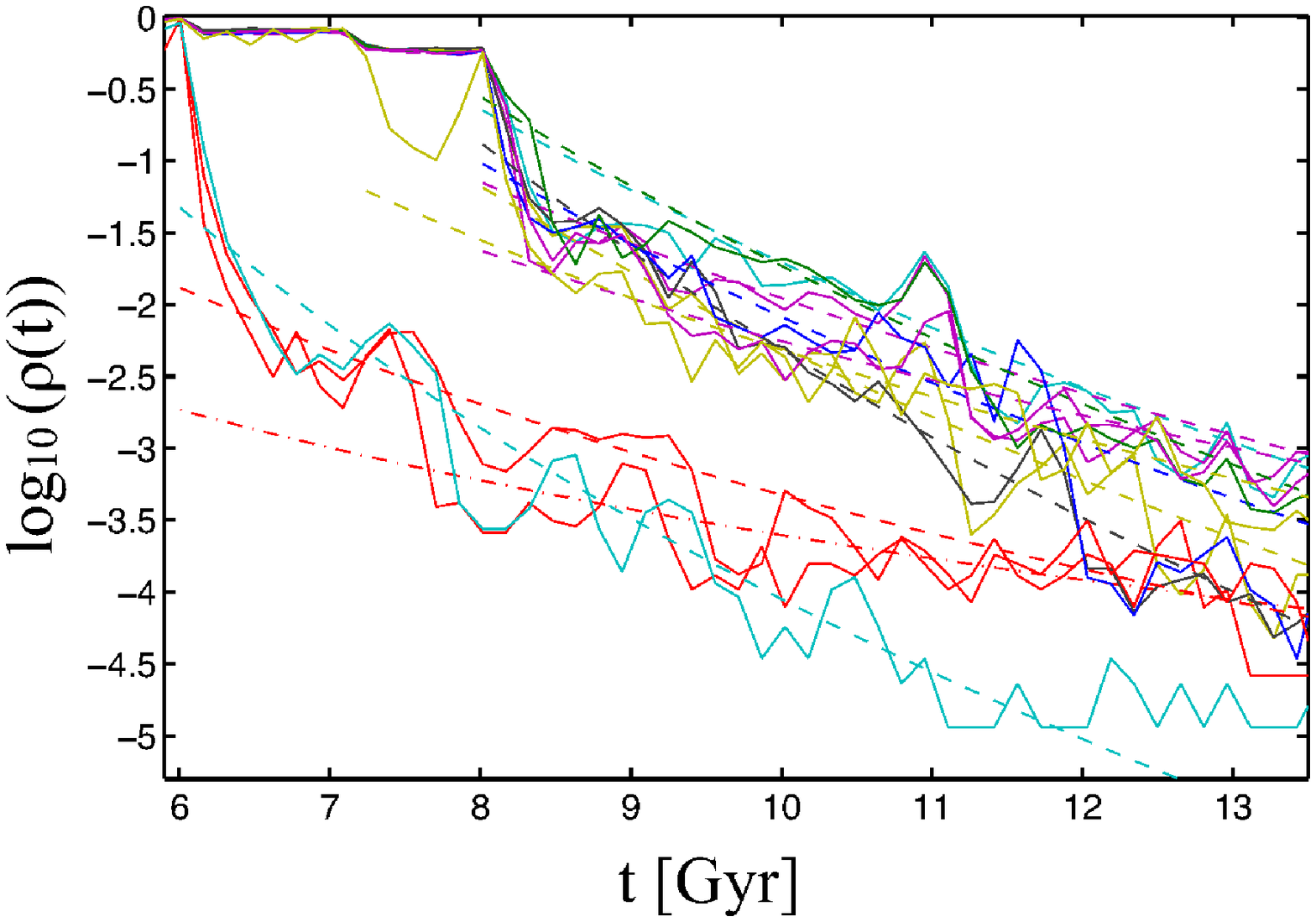}
\caption{Time evolution  of the density of streams  originating in two
  different  satellites that  contribute,  at redshift  $z=0$, to  the
  solar neighbourhood  sphere obtained  from halo Aq-A-2.   Each panel
  corresponds  to a  different satellite.  The dashed  lines  show the
  power-law  fits  applied  to  the  particles  identified  as  stream
  members, as  explained in Section~\ref{sec:unresolved},  whereas the
  dashed-dotted lines correspond to the remaining particles.}
\label{fig:dens_evol}
\end{figure*}

As  in Section~\ref{sec:resolved_subs},  we place  a sphere  of  4 kpc
radius around  each particle  at $t  = t_{\rm form}$  and tag  all the
surrounding neighbours. We track these particles forward in time until
redshift $z=0$ and those neighbours  that depart from our particles by
more than $r_{\rm stream}$ at  any time are discarded.  We compute the
local spatial density of the selected particle, $\rho(t)$, by counting
the number of neighbours within  a distance $r_{\rm dens} = 0.5 \times
r_{\rm  apo}$.    In  Figure~\ref{fig:dens_evol}  we   show  the  time
evolution  of the  spatial  density of  particles  from two  different
satellites that contribute to  the ``solar neighbourhood'' of the halo
Aq-A-2.  From this figure we can appreciate that the rate at which the
spatial  density  of a  stream  decreases  with  time varies  strongly
(likely  as a  consequence  of  the different  orbits).   We can  also
observe that in  many cases, when a particle  becomes unbound from its
parent  satellite, the  density of  the newly  formed  stream exhibits
initially a  very rapid and  quasi-exponential decrease.  As  shown by
\citet{g10} \citep[see  also][]{hg07}, this initial  transient is also
present  in   regular  orbits  and   does  not  necessarily   imply  a
manifestation  of chaotic  behaviour.   It is  therefore necessary  to
follow  the evolution  of the  stream's density  for long  time scales
(much longer than a crossing time) to disentangle regular from chaotic
behaviour.

Our approach to characterizing the  type of mixing consists in fitting a
power-law function  to the  time evolution of  the local density  of a
stream,
\begin{equation}
\label{eq:rho_t}
\hat{\rho}(t) = \alpha t^{-n},
\end{equation}
and  determining  the value  of  $n$ for  each  stream.   As shown  by
\citet{vol08} \citep[see  also][]{hw} for a stream on  a regular orbit,
$n=1$,  2  or  3  depending  on the  number  of  fundamental  orbital
frequencies.   Although we  know that  the density  of a  stream  on a
chaotic orbit does  not evolve as a power law  in time, we nonetheless
fit  the functional  form  given by  Eq.~(\ref{eq:rho_t}), and  expect
larger values  of $n$ than for  the regular case.  We  warn the reader
that,  given  our  simple  approach, the  resulting  distributions  of
$n$-values  should be  considered as  an estimate  of the  rate  at which
diffusion  is  acting  in  a  given  volume,  rather  than  a  precise
characterization of the underlying phase-space structure.

Figure~\ref{fig:dens_evol} shows the results of applying such a fit to
the stream densities as a function of time.  Since we are interested
in the behaviour of the density on long time scales, we do not include
the initial quasi-exponential transient in the fits. We proceed as
follows:
\begin{itemize}
\item We define  the formation time of a  stream, $t_{\rm stream}$, as
  the time  when its density decreases  from one snapshot  to the next
  one by $50\%$.
\item Starting from the  snapshot associated with $t_{\rm stream}$, we
  iteratively  fit ten  times  a power-law  function  to the  stream's
  density  by increasing  in each  iteration  the starting  time
    snapshot by snapshot. The fits are always normalized to the value
  of the density at the  starting point.  Furthermore, each data point
  is weighted according to the number of particles used to measure the
  corresponding density.
\item  We estimate the  goodness of  each fit  by computing the root mean
  squared error, $RMSE$, defined as:
\begin{equation}
\displaystyle
RMSE = \sqrt{\dfrac{\sum_{i=1}^{m}\left(\hat{\rho}(t_{i}) - \rho(t_{i})\right)^{2}}{m-2}}
\end{equation}
where $\rho(t_{i})$ is the  density estimated from the particle count
and $m$ is the number of data points used in the fit.
\item Of all these experiments, we only keep the value of $n$ obtained
  from the fit where the $RMSE$ is smallest.
\end{itemize}

\begin{figure*}
\includegraphics[width=180mm,clip]{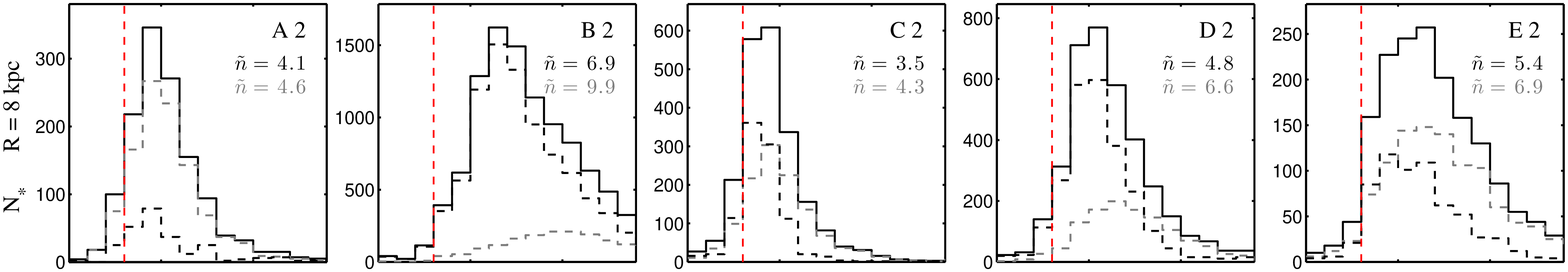}
\\
\includegraphics[width=180mm,clip]{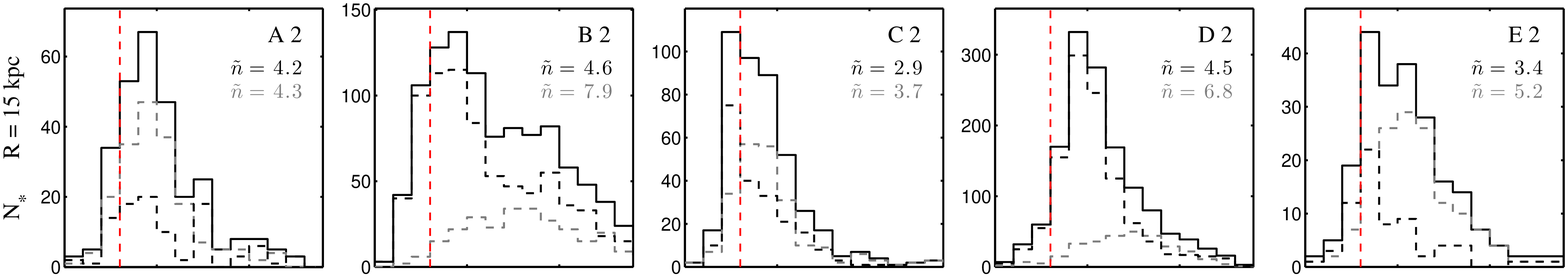}
\\
\includegraphics[width=180mm,clip]{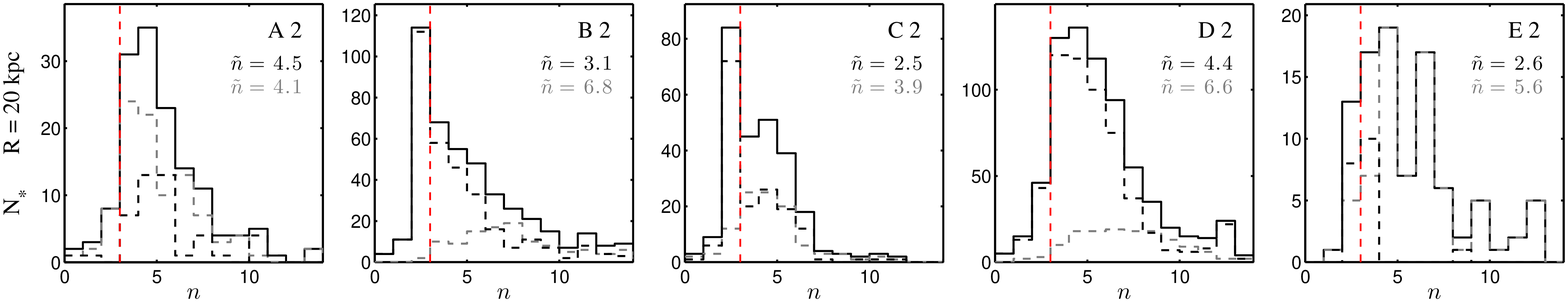}
\caption{The   black  solid  histograms   show  the   distribution  of
  $n$-values found for particles in ``solar neighbourhood'' spheres of
  2.5 kpc  radius.  Each column corresponds to  a different satellite.
  Different   rows  corresponds  to   volumes  centred   at  different
  galactocentric  radius.    The  black  (respectively   grey)  dashed
  histograms  correspond to  those particles  that  are (respectively,
  not)  associated  to resolved  stellar  streams.   The vertical  red
  dashed line is  located at $n=3$. For regular orbits,  $n \leq 3$ is
  expected.  Each panel shows  the median values $\tilde{n}$, obtained
  from the  distributions for particles in resolved  and in unresolved
  streams.}
\label{fig:histo}
\end{figure*}

The results of applying this procedure to all the particles inside 2.5
kpc   radius    spheres   of    different   haloes   are    shown   in
Figure~\ref{fig:histo}.  We explore spheres located at three different
galactocentric  radii, namely 8  , 15  and 20  kpc.  The  black solid
histograms show  the distribution of $n$-values for  all the particles
in each  ``solar neighbourhood''  sphere.  At ${\rm  R} = 8$  kpc (top
panels), with the  exception of haloes Aq-B-2 and  E-2, we find narrow
distributions  peaking at  values  slightly larger  than $n=3$.   Note,
however, that as  we move outward from the  galactic centre, the peak
of  all distributions  tend to  shift  towards values  much closer  to
$n=3$.   A clear  example is  halo Aq-B-2,  where the peak  of the
distribution shifts  from $n  \approx 7$ at  ${\rm R}  = 8$ kpc  to $n
\lessapprox 3$ at ${\rm R} = 20$ kpc (bottom panel).

Figure~\ref{fig:histo}  also  shows  in  black dashed  histograms  the
distribution $n$-values for those particles associated with ``resolved''
stellar streams,  while the  grey-dashed histograms correspond  to the
remaining particles.   From this figure  we can observe  that although
the two  distributions overlap significantly, there is  a clear offset
between  them.   In each  panel  we show  the  median  values of  $n$,
$\tilde{n}$.  Stellar  particles in streams (which are  more likely to
be on regular orbits) tend to  have the smallest values of $n$ and, in
general, their  distribution peaks at a  value much closer to  $n = 3$
than their  unresolved counterparts.  Let  us recall that this  is the
value of $n$  expected for the dependence of density  on time for
regular orbits  in a three dimensional  time independent gravitational
potential \citep[see, e.g.][]{hw,vol08}.

It is interesting to compare the $n$-distributions obtained from haloes
Aq-A-2 and  B-2 at  ${\rm R}=8$ kpc.   As previously  discussed, while
halo B-2 has  the largest number of streams  and the greatest fraction
of accreted particles associated with  them, the opposite is found for
halo Aq-A-2.   The value  of $\tilde{n} \approx  7$ obtained  from the
resolved  component in  halo  B-2 suggests  that  these particles  are
undergoing chaotic  mixing.  Nonetheless, thanks to  the high particle
resolution,  substructure  can be  efficiently  identified.  Note  the
large difference  between the medians  obtained from the  resolved and
unresolved  components.   For  halo   Aq-A-2  ,  the  distribution  of
$n$-values  associated with  the two  components show  a  very similar
value  of $\tilde{n}  \approx  4.3$, considerably  closer  to what  is
expected for regular orbits.  This suggests that a significant fraction
of  the particles  that have  not been  associated with  streams  are on
(nearly)  regular orbits, and  that because  of the  limited numerical
resolution, they are not found in (massive) streams.

Interestingly, we find that, on average, particles in resolved streams
in  Aq-A-2  were released  from  their  parent satellite\footnote{  As
  measured by  the stream's formation time, $t_{\rm  stream}$} 2.5 Gyr
later  and  have  more  extended  orbits than  the  particles  in  the
unresolved component.  The average  apocentric distances for these two
subsets are  32 and 12  kpc, respectively.  Thus, the  shorter orbital
periods  and longer  times  since release  partially  explain why  the
latter are not  found to be part of (massive)  streams.  A lower limit
to  the fraction of  accreted particles  in resolved  stellar streams,
corrected by  resolution effects, can  be obtained by  quantifying the
fraction  of particles  in  the unresolved  component with  $n$-values
smaller  than the  median of  the resolved  component, $\tilde{n}_{\rm
  res}$.  We find in halo Aq-A-2 a total of 411 particles that satisfy
this condition.  In addition to  the 283 particles associated with the
resolved component, we  estimate that at least $50\%$  of all accreted
particles  in  this  halo  should   be  part  of  a  resolved  stellar
stream. Furthermore, the corresponding stellar mass fraction should at
least be as large as $65.7\%$.

In general,  particles in resolved  streams were released  later (than
those in  the unresolved component) for  all haloes. We  find for haloes
Aq-B-2, C-2, D-2 and E-2 a time  difference of 1.9, 1, 2.2, and 2 Gyr,
respectively.  However,  differences in the  mean apocentric distances
are found only in halo Aq-A-2.   As for halo Aq-A-2, we can obtain for
the remaining  haloes an estimate for  the lower limit  of particles in
resolved stellar  streams corrected by resolution effects.   We find a
total  of $(380,348,268,350)$ particles  with $n$-values  smaller than
$\tilde{n}_{\rm    res}$   in    haloes   (B-2,    C-2,    D-2,   E-2),
respectively. Thus,  we estimate  that the corresponding  fractions of
particles and stellar mass in  resolved streams for these haloes should
be, at least, as large  as $(55\%, 75\%, 62\%,84\%)$ and $(96\%, 80\%,
91\%, 84\%)$, respectively.

\section{Summary}
\label{sec:conclu}

In this work we have characterized the phase-space distribution as
well as the time evolution of debris for stars found inside ``solar
neighbourhood'' volumes.  Our analysis is based on simulations of
the formation of galactic stellar haloes, as described by
\citet{cooper}.  The haloes were obtained by combining the very high
resolution fully cosmological $\Lambda$CDM simulations of the {\it
Aquarius} project with a semi-analytic model of galaxy formation.

We find that, for haloes  Aq-A-2, -C-2 and -D-2, the measured velocity
ellipsoid at 8 kpc on the major  axis of the halo is in good agreement
with   the   estimate   for    the   local   Galactic   stellar   halo
\citep{chibabeers}.  Similarly,  for all haloes but  Aq-B-2, the local
stellar  halo average  density  on  the major  axis  is in  reasonable
agreement  with the  observed value  \citep{fuchs}.  However,  we note
that if  the contribution of the  disc and bulge were  to be included,
the resulting velocity  dispersions as well as the  local stellar halo
average density would be  significantly increased.  This could suggest
that these dark haloes are too massive to host the Milky Way \citep[in
agreement with][]{giusy,s07,mcs,wang,vera13}.   On the other  hand, if
located in a prolate low mass  halo such as Aq-B-2, the measured local
stellar density would imply that the Sun should be on the intermediate
or minor  axis, i.e.   the Galactic disc's  angular momentum  would be
aligned with  the major  axis of the  dark halo \citep{ah04}.   On the
minor  axis,  haloes  B  and  E have  velocity  ellipsoids  of  smaller
amplitude  and  hence could  perhaps  be  made  consistent with  those
observed near  the Sun, if the  effects of the  contraction induced by
the disc were taken into account.  However, for haloes A, C, and D, the
dispersions on the minor axis are  larger than those on the major axis
at the equivalent of the solar circle.

In agreement with previous work, we find that $90\%$ of the stellar
mass  enclosed in our  ``solar neighbourhoods'' comes, in  all cases,
from 3  to 5  significant contributors. These  ``solar neighbourhood''
volumes contain a large amount  of substructure in the form of stellar
streams.   In the five  analysed stellar  haloes, substructure  can be
easily identified as kinematically cold structures, as well as in well
defined  lumps  of stellar  particles  in  spaces of  pseudo-conserved
quantities.

By applying a  simple algorithm to follow the  time evolution of local
densities  in  the  neighbourhood  of  reference  particles,  we  have
quantified the  number of  (massive) streams that  can be  resolved in
solar    neighborhood-like    volumes.    In   haloes    where    solar
neighbourhood-like volumes  are best resolved we find  that the number
of resolved  streams is  in very good  agreement with  the predictions
given by \citet{hw}.  As we  move outward from the galactic centre the
number of streams decreases  as a power-law, in a  similar fashion to the
density profiles.

To explore  whether the  halo-to-halo scatter in  the total  number of
resolved  streams is  due to  poor particle  resolution or  to chaotic
mixing, we have  estimated the rate at which  local (stream) densities
decrease with time.  Interestingly,  we find that in our best-resolved
solar   neighbourhood-like   volume  (Aq-B-2)   a   large  amount   of
substructure  can  be identified,  even  in  the  presence of  chaotic
mixing.  In this halo, up to  $82\%$ of the star particles (and $92\%$
of the stellar mass) can be associated with a resolved stream.  On the
other  hand,  our  poorest-resolved  solar  neighbourhood-like  volume
(Aq-A-2)  shows a  smaller  amount of  substructure but  significantly
longer mixing  time-scales (a rate that  is close to  what is expected
from regular orbits).  In this  case, only $20\%$ of the accreted star
particles (and $31\%$  of the accreted stellar mass)  can be linked to
streams.   A comparison  of  the orbital  properties  of resolved  and
unresolved  stream-populations shows  that, in  general,  particles in
resolved streams  are released later and have  longer orbital periods.
Moreover, diffusion in both  populations occurs at very similar rates.
Thus,  this  analysis  suggests   that  our  strongest  limitation  to
quantifying substructure is mass  resolution rather than diffusion due
to chaotic mixing. We have estimated  a lower limit to the fraction of
accreted  particles  in resolved  stellar  streams  by correcting  for
resolution effects.  We  find that this fraction should  be, at least,
as large as $50\%$ by number and $65\%$ by stellar mass in all haloes.

The  results  presented in  this  work  suggest  that the  phase-space
structure  in  the  vicinity  of  the  Sun should  be  quite  rich  in
substructure. Recall that our analysis is based on fully cosmological
simulations   and  therefore   chaos  and   violent  changes   in  the
gravitational potential that are so characteristic of cold dark matter
model, are naturally included. 

With  the imminent  launch  of the  astrometric  satellite {\it  Gaia}
\citep{perry}, we will be able for the first time to robustly quantify
the  amount of  substructure in  the form  of stellar  streams present
within $\approx 2$  kpc from the Sun.  This  mission from the European
Space Agency  will provide accurate measurements  of positions, proper
motions, parallaxes and radial  velocities of an unprecedentedly large
number        of         stars\footnote{for        details        see:
  http://www.rssd.esa.int/gaia}.   In  addition,  recent  and  ongoing
surveys  such  as   SEGUE  \citep{segue},  RAVE  \citep{rave},  LAMOST
\citep{lamost},      HERMES      \citep{hermes}      and      Gaia-ESO
\citep{2012Msngr.147...25G}  will allow  us to  complement  these {\it
  Gaia}  observations.    A  direct   comparison  of  the   degree  of
substructure  found in  the  Solar Neighbourhood  with  that found  in
numerical  models  such as  those  presented  here  will allow  us  to
establish directly, for example, how much of the stellar halo has been
accreted and how much has been formed in-situ.

\section*{Acknowledgments}
The authors would like to thank the anonymous referee for his/her very
useful  comments and suggestions  that helped  improve the  clarity of
this paper.  FAG would  like to thank  Brian W.  O'Shea  for providing
useful  comments  and  suggestions  and Monica  Valluri  for  valuable
discussions.    FAG  was   supported   through  the   NSF  Office   of
Cyberinfrastructure by  grant PHY-0941373,  and by the  Michigan State
University Institute for Cyber-Enabled  Research (iCER). This work was
initiated   thanks   to  financial   support   from  the   Netherlands
Organisation for  Scientific Research (NWO)  through a VIDI  grant. AH
acknowledges   the  European  Research   Council  for   ERC-StG  grant
GALACTICA-240271.  APC  acknowledges support from  the Natural Science
Foundation of China, grant No. 11250110509.

\bibliographystyle{mn2e}
\bibliography{aquarius}

\label{lastpage}

\end{document}